%% file: paper.tex
\documentclass[prd,preprintnumbers,nofootinbib,superscriptaddress,tightenlines,twocolumn,notitlepage,noeprint]{revtex4-2}

\input{texdefinitions_papers.tex}
\usepackage{booktabs}
\usepackage{tabularx}
\usepackage{array}

\titlespacing{\section}{0pt}{10pt plus 1pt minus 1pt}{2pt plus 1pt minus 1pt}
\titlespacing{\subsection}{0pt}{2pt plus 0pt minus 0pt}{2pt plus 0pt minus 0pt}
\titlespacing{\subsubsection}{0pt}{2pt plus 0pt minus 0pt}{2pt plus 0pt minus 0pt}

\begin{document}

\title{Event-Level QCD Inference Framework for Quark-Gluon Imaging}

\author{Patrick Barry}
%\email{barry@anl.gov}
\address{Physics Division, Argonne National Laboratory, Lemont, IL 60439, USA}
\author{Pi-Yueh Chuang}
%\email{pchuang@anl.gov}
\address{Department of Computer Science, Virginia Tech, Blacksburg, VA 24061, USA}
\address{Mathematics and Computer Science Division, Argonne National Laboratory, Lemont, IL 60439, USA}

\author{Ian Clo\"{e}t}
%\email{icloet@anl.gov}
\address{Physics Division, Argonne National Laboratory, Lemont, IL 60439, USA}

\author{Emil Constantinescu}
%\email{emconsta@anl.gov}
\address{Mathematics and Computer Science Division, Argonne National Laboratory, Lemont, IL 60439, USA}

\author{Arkaprabha Ganguli}
%\email{aganguli@anl.gov}
\address{Mathematics and Computer Science Division, Argonne National Laboratory, Lemont, IL 60439, USA}

\author{Chao Peng}
%\email{cpeng@anl.gov}
\address{Physics Division, Argonne National Laboratory, Lemont, IL 60439, USA}

\begin{abstract}
We introduce and demonstrate an event-level analysis framework for quark-gluon imaging. For a first application we use it for the inference of parton distribution functions from synthetic deep inelastic scattering data. This framework removes the need for unfolding of detector effects and the binning of events, and therefore eliminates two key sources of information loss. We contrast this event-level framework with the traditional histogram approach by performing a closure test for parton distribution functions from event data obtained from a known ground truth. In this study we assume a perfect detector, which makes unfolding straightforward. The elimination of binning in the event-level framework is demonstrated to have important benefits over the traditional histogram approach, and performs better in the closure test, particularly for a smaller number of events. For example, defining a mean-squared error distance metric, we find that the event-level framework performs around 35\% better than the traditional approach for a moderate number of events. The benefits of an event-level framework should increase for inference associated with 3D quark-gluon imaging, because these differential cross sections are of higher dimension and the comparative number of measured events is significantly reduced. 
\end{abstract}

\maketitle
%-------------------------------------------------------------------------------
%-------------------------------------------------------------------------------
\section{INTRODUCTION}
\looseness=-1
The multidimensional imaging of the quarks and gluons that comprise protons and nuclei is a central priority for facilities such as Jefferson Lab~\cite{Dudek:2012vr,Arrington:2021alx} and a key scientific driver for the forthcoming Electron Ion Collider (EIC)~\cite{Accardi:2012qut,AbdulKhalek:2021gbh}. This imaging is not direct, however, because of quark-gluon confinement~\cite{Ellis:1996mzs} in Quantum Chromodynamics (QCD). Instead, imaging is achieved by solving inverse problems on data from certain scattering processes using QCD factorization theorems~\cite{Collins:1989gx,Collins:2011zzd}, which relate observable cross sections to quantum correlation functions (QCFs). These QCFs encode information on the internal quark-gluon dynamics of hadrons and nuclei, where important examples for quark-gluon imaging include parton distribution functions (PDFs), transverse momentum-dependent distributions (TMDs), and generalized parton distributions (GPDs). 

At a fundamental level, the experimental information for quark-gluon imaging is provided by collections of measured scattering events and total cross sections from processes that are sensitive to these QCFs, such as deep inelastic scattering (DIS), semi-inclusive deep inelastic scattering (SIDIS), and deeply virtual exclusive scattering (DVES). A scattering event is simply a collection of momentum vectors, with corresponding particle type and state (e.g., spin), for the incoming and measured outgoing particles in a scattering process. For DIS this is just the momentum vectors and spin state for the incoming lepton (usually an electron) and target, and the outgoing momentum vector for the scattered lepton, for SIDIS an event also includes the momentum vector of an outgoing hadron (e.g., a pion or kaon), and for DVES processes the momentum vector of the outgoing target can be included in the definition of an event because the target remains intact.

The standard or traditional approach for the QCD analysis associated with quark-gluon imaging does not operate directly on these events. Instead, measured scattering data are first processed through unfolding~\cite{DAgostini:1994fjx,Hocker:1995kb,Andreassen:2019cjw,Arratia:2021otl,Chan:2023tbf,Shmakov:2023kjj,Glazov:2017vni,Schofbeck:2024zjo,Benato:2025rgo,Desai:2025mpy,Milton:2025mug,Ore:2026qgp,Pate:2026swu} procedures to correct for detector effects and then histogrammed or binned over the physical or true kinematic phase space of the process. This procedure gives a histogram approximation to the underlying true differential cross section, where the statistical uncertainty for each bin can be determined via bootstrap methods. QCFs are subsequently extracted through global analyses of these derived bin-level observables~\cite{NNPDF:2021njg,Barry:2025glq,Bacchetta:2025ara,Guo:2025muf}. 

The traditional histogram method has proven successful over many decades~\cite{Bloom:1969kc,Breidenbach:1969kd,Miller:1971qb,Bailey:2020ooq,Hou:2019efy,PDF4LHCWorkingGroup:2022cjn,Cocuzza:2021cbi}, however, this workflow introduces two main sources of information loss associated with the unfolding and binning procedures. The unfolding process, which maps events from the measured phase space to the true phase space is an ill-posed inverse problem~\cite{Nachman:2025_unfoldAI,Canelli:2025ybb} and necessarily involves approximations, while binning replaces individual events with averages over finite regions of the true phase space. This loss of information is likely moderate for the inference of PDFs from DIS observables, where there are abundant data and the differential cross sections are of low dimension. However, these effects will become increasingly severe for processes relevant to 3D imaging of quarks and gluon\emdash such as TMD- and GPD-sensitive reactions\emdash where differential cross sections are of much higher dimension and available event data are comparatively limited.

To address this loss of information, we propose a new {\it event-level} analysis framework for the inference of QCFs directly from the measured scattering events and total cross sections (or equivalently integrated luminosities). This framework maintains the maximum amount of information in the experimental data by removing the need for the unfolding and binning procedures. In this event-level workflow, detector effects are \textit{folded} into theory-generated true events, so that theory and experiment can be directly compared at the measured event level without the need for binning.

As a first demonstration of this framework we perform an event-level inference for unpolarized PDFs from synthetic DIS events at EIC kinematics, and contrast these results with the traditional histogram approach using the same data. For this initial study we assume a perfect detector, which makes unfolding trivial, and leave comparison between event-level and binning for the higher dimensional differential cross sections associated with TMDs and GPDs for a future study. 

The structure of this manuscript is as follows: Sec.~\ref{sec:DIS} discusses the DIS process, PDFs, and DIS events; Sec.~\ref{sec:eventlevel} introduces the event-level analysis framework with a focus on DIS; Sec.~\ref{sec:binlevel} discusses the traditional histogram approach for DIS; Sec.~\ref{sec:results} contrasts results obtained using the new event-level analysis framework with those using the histogram approach for PDFs extracted from ground truth synthetic DIS events; and Sec.~\ref{sec:conclusions} provides our conclusions and an outlook.

%-------------------------------------------------------------------------------
%-------------------------------------------------------------------------------
\section{THE DIS PROCESS AND DIS EVENTS\label{sec:DIS}}
\begin{figure}
\centering\includegraphics[width=\columnwidth]{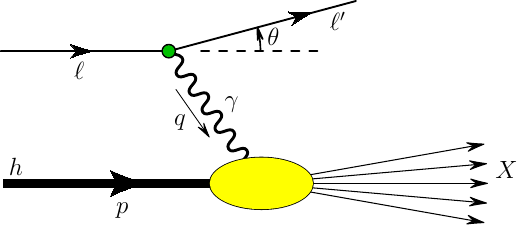}
\caption{Illustration of the DIS process in the one-photon exchange approximation. The quantities $\ell $ and $\ell'$ are the initial and final lepton momenta, $h$ represents the type of target with momentum $p$, $q = \ell'-\ell$ is the momentum transferred to the target by the virtual photon $\gamma$, and $X$ is the hadronic final state. The angle $\theta$ between the incoming and outgoing lepton is defined in the target rest frame.}
\label{fig:dis}
\end{figure}

The DIS process in the one-photon exchange approximation is illustrated in Fig.~\ref{fig:dis}, and is represented by $l(\ell) + h(p) \to l(\ell') + X$, where $\ell $ and $\ell'$ are the initial and final lepton momenta, $h$ represents the type of target (e.g., proton or nucleus) with momentum $p$, $q = \ell'-\ell$ is the momentum transferred to the target by the virtual photon $\gamma$, and $X$ represents the multi-particle final state which is not measured in DIS. The form of the differential cross section for this process is constrained by Lorentz invariance, parity, and time reversal symmetries, and for an unpolarized spin-half target reads~\cite{Blumlein:1996vs,Workman:2022ynf}\footnote{The differential cross section $\dd \sigma^h / \dd x \dd Q^2$ can be interpreted as an unnormalized probability density over $(x,Q^2)$ for the DIS process at fixed center-of-mass energy $\sqrt{s}$. Equation~\eqref{eq:probdis} normalizes it by the total cross section.}
\begin{align}
\label{eq:DISxsec}
\sigma^h(x,Q^2) &\coloneq \frac{\dd \sigma^h}{\dd x \dd Q^2} = \frac{2\pi \alpha^2_e}{x\,Q^4} \\
&\hspace*{-13mm}
\times\left[\left[1+(1-y)^2 - 2\,x^2 y^2 m_h^2/Q^2\right]F_2^h(x,Q^2) - y^2\,F_L^h(x,Q^2)\right], \nonumber
\end{align}
where $\alpha_e \simeq 1/137$ is the fine structure constant, $F_{2,L}^h$ are the structure functions which characterize the interaction of the virtual photon with the target and its remnants where $F_L \coloneq F_2 - 2\,x\,F_1$~\cite{Blumlein:1996vs,Vermaseren:2005qc}, $m_h$ the mass of the target, $Q^2 \coloneq -q^2$,  $x \coloneq Q^2/(2\,p\cdot q)$ is the Bjorken scaling variable, $y \coloneq p\cdot q/p\cdot \ell = Q^2/[x\,(s-m_h^2)]$ is the fraction of energy transferred to the target in its rest frame, where $s \coloneq (\ell+p)^2$ is the lepton-target center-of-mass energy squared, and $0 < x,y < 1$. Another important kinematic variable is $W^2 \coloneq (p+q)^2 = m_h^2 + Q^2[1/x-1]$, where $W$ is interpreted as the mass of the hadronic final state $X$, which for DIS kinematics is often set to $W^2 \gtrsim 10\,$GeV$^2$~\cite{NNPDF:2021njg,Moffat:2021dji}, however, values of $W^2 \gtrsim 3\,$GeV$^2$ are also used~\cite{Cocuzza:2021cbi}.

The physical phase space for a DIS experiment is determined by characteristics of the experiment (e.g. beam energy) and the kinematic cuts imposed to ensure the validity of the factorization theorems. We work with typical EIC kinematics and take $\sqrt{s} = 140\,$GeV, then the kinematic constraint $0 < y < 1$, and the imposed cuts $Q^2_{\rm min} = m_c^2 = (1.28\,{\rm GeV})^2$ and $W^2_{\rm min} = 10\,$GeV$^2$ defines the physical phase space in terms of $x$ and $Q^2$. These constraints also give the relations $Q^2_{\rm max} = s - W^2_{\rm min}$, $x_{\rm min} = Q_{\rm min}^2/(s-m_h^2)$, and $x_{\rm max} = Q_{\rm max}^2/(s-m_h^2)$.

%---------------------------------------------------------------------
%---------------------------------------------------------------------

The structure functions $F_2^h$ and $F_L^h$ are target dependent observables, and very difficult to calculate directly from QCD, so knowledge of these functions has been extracted from DIS experiments. For DIS kinematics, QCD factorization implies that the structure functions can be expressed as
\begin{align}
F^h_i(x,Q^2) &= \nonumber \\
&\hspace*{-7mm}
\sum_q \lf[C^q_i(y) \otimes q^h(z,Q^2) + C^g_i(y) \otimes g^h(z,Q^2)\rg] + \ldots, 
\label{eq:F2L}
\end{align}
where $i=(2,L)$, the convolution is given by $b(y)\otimes c(z) \coloneq \iint_0^1\dd y \dd z\ \delta(x-yz)\  b(y)\,c(z)$, 
the sum is over active quark flavors ($q=u,\bar{u}, d,\bar{d}, \ldots$), the $C_i$ are perturbatively calculable target independent hard coefficient functions, and the ellipsis represents various corrections to the factorization formula, such as power corrections of the form $\Lambda_{\rm QCD}/Q^2$ and $M^2/Q^2$~\cite{Collins:2011zzd}. The quark and gluon distribution functions, $f^h = (q^h,\,g^h)$, of the target $h$ are directly related to aspects of quark-gluon dynamics associated with internal structure of $h$. Extracting these PDFs from data therefore provides important insights to QCD and how quarks and gluons bind together to form hadrons and nuclei~\cite{Barry:2018ort,Aubert:1983xm,Cocuzza:2021rfn}.

To contrast our event-level analysis framework with traditional histogram methods, we need to generate synthetic event data from a ground truth differential cross section. This is achieved by first defining ground truth forms for the PDFs $q^h$ and $g^h$. Using Eq.~\eqref{eq:DISxsec} and Eq.~\eqref{eq:F2L} with next-to-leading coefficient functions gives the two structure functions and the ground truth differential cross section. For the ground truth, and forthcoming inference, the PDFs take the general form
\begin{align}
f(x,Q_0^2;\boldsymbol{\theta}) = \mathcal{N}\, x^\alpha (1-x)^\beta\,
(1 + \gamma \sqrt{x} + \delta\, x),
\label{eq:PDFparam}
\end{align}
where the parameters $\boldsymbol{\theta} = (\mathcal{N},\alpha,\beta,\gamma,\delta)$ are PDF-type and target dependent, and we take $Q_0^2 = Q^2_{\rm min}$. For the ground truth we assume both a proton and a neutron beam at the EIC (thereby ignoring nuclear effects which are not central to this study), and assume three active quark flavors, $q = u,\bar{u}, d,\bar{d}, s,\bar{s}$, together with an active gluon PDF. For the ground truth we base our parameters on a typical central replica for the JAM Collaboration PDF analysis of Ref.~\cite{Anderson:2024evk}, and assume the standard charge symmetry relations between proton and neutron PDFs, that is, $u \coloneq u^p = d^n$, $d \coloneq d^p = u^n$, with analogous relations for $\bar{u}$ and $\bar{d}$, and finally $s \coloneq s^p = s^n = \bar{s}^n = \bar{s}^p$. These PDFs satisfy the $Q^2$ independent baryon number and momentum sum rules:
\begin{align}
& \int_0^1 \dd x\ u^-(x) = 2, \qquad \int_0^1 \dd x\ d^-(x) = 1, \\
& \int_0^1 \dd x\ x \left[u(x) + \bar{u}(x) + d(x) + \bar{d}(x) + \ldots + g(x) \right] = 1,
\label{e:sumrules}
\end{align}
where the minus-type or valence PDFs are given by $q^- = q_v = q - \bar{q}$. These PDFs evolve from $Q_0^2$ to $Q^2$ according to the DGLAP evolution equations~\cite{Altarelli:1977zs},  using next-to-leading order (NLO) splitting functions in the zero-mass variable flavor number scheme. This process gives the $Q^2$ dependence of the structure functions and thereby the DIS differential cross section, and also dynamically generates charm and bottom quark PDFs at their mass thresholds.

The next step is to obtain the synthetic events. This is possible because the differential cross section of Eq.~\eqref{eq:DISxsec} is a non-negative function of the variables $x$ and $Q^2$, which when integrated over the physical phase space gives the total cross-section $\sigma$. The total cross section is given empirically by the formula
\begin{align}
n = \sigma\,\mathcal{L},
\label{eq:eventscrosssection}
\end{align}
where $n$ is the number of observed events and $\mathcal{L}$ is the total/integrated luminosity, which is determined by many experimental setup parameters, such as characteristics of the beam and target, run time, and detector coverage~\cite{wille2000physics,Herr:2003em}. This makes clear that in scattering experiments the fundamental units of information are provided by the total cross section and the scattering events; everything else is derived from these observables, including the differential cross section and structure functions. For DIS, the measured events are a set of size $n$ where each element is a $(x,Q^2)$ pair. Then the true differential cross section is inferred from $\sigma$ and the distribution of events over either the physical (unfolding) or measured (folding) phase space. Therefore, events are not {\it function} samples from the differential cross section but instead random $(x_i,Q^2_i)$ {\it density} samples distributed, in the physical phase space $\Omega$, according to the probability distribution 
\begin{align}
\hat{\sigma}(x,Q^2) = \frac{1}{\sigma}\, \sigma(x,Q^2)\Big|_{\Omega}, \quad \text{where}\quad
\sigma = \int_{\Omega} \dd x\dd Q^2\ \sigma(x,Q^2).
\label{eq:probdis}
\end{align}
Therefore, the synthetic events are obtained by randomly sampling $\hat{\sigma}(x,Q^2)$ over $\Omega$, where we use Monte Carlo rejection sampling~\cite{devroye_chapter_2006}. 

%-------------------------------------------------------------------------------
%-------------------------------------------------------------------------------
\section{EVENT-LEVEL ANALYSIS FOR DIS\label{sec:eventlevel}}
\begin{figure}
\centering\includegraphics[width=\columnwidth]{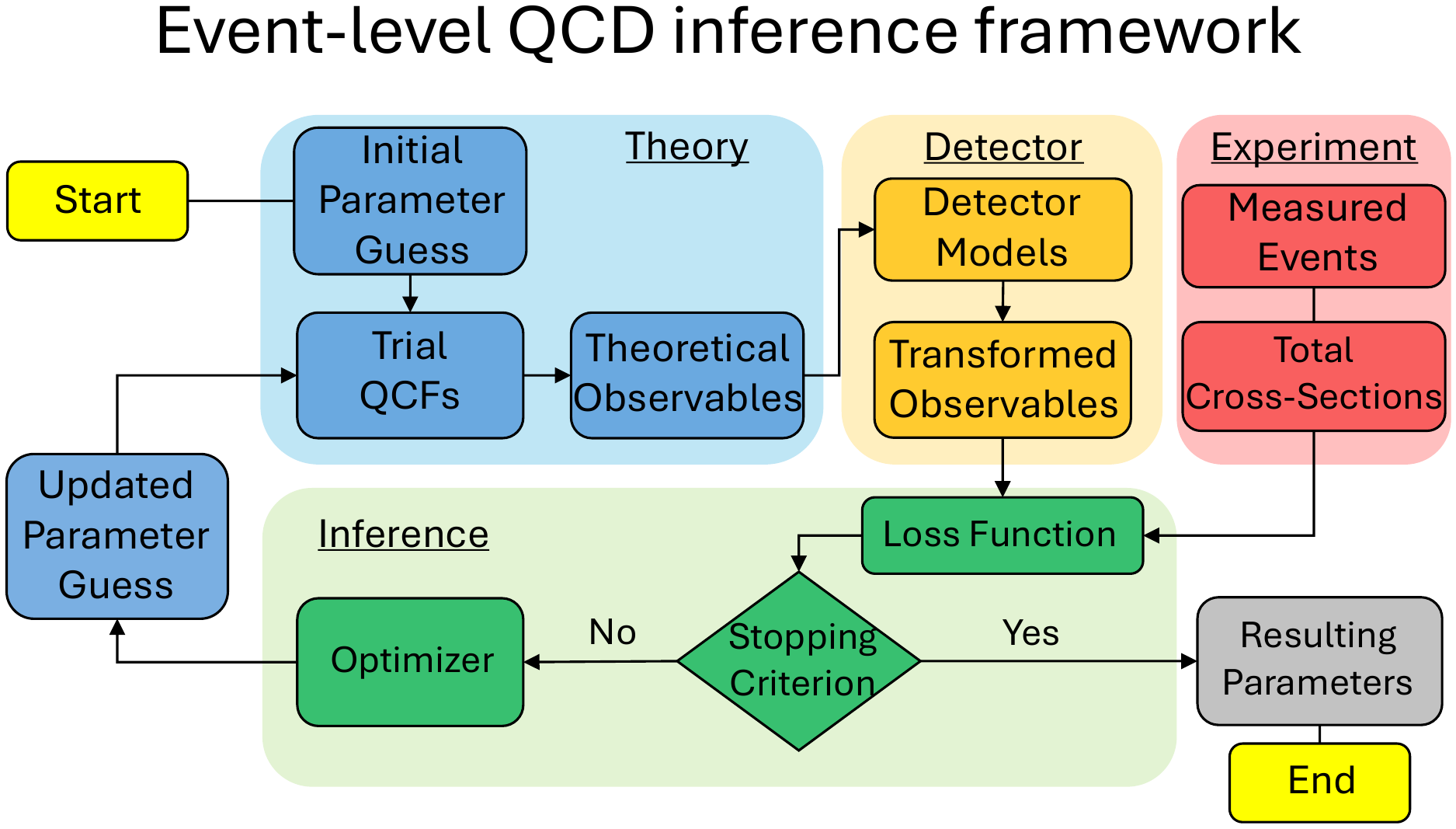}
\caption{Schematic of the event-level analysis framework. It begins with an initial guess for the parameters for the QCFs, which can be parameters for parametric functions, neural network weights, etc. These parameters define trial QCFs which form the true differential cross sections. In Method I, these differential cross sections are sampled and these true events pass through a detector model becoming synthetic measured events. In Method II, the true differential cross sections past through the detector model to give the differential cross sections in the measured phase space. Finally, the loss function compares the synthetic and measured observables, and this process repeats until a minimum is found for the parameters or other stopping criteria are met.}
\label{fig:workflow}
\end{figure}

An event-level analysis framework for inference of QCFs from experimental data, means comparing theory directly with the measured scattering events and associated total cross sections. Broadly, we have considered two approaches to perform an event-level analysis for QCFs:
\begin{itemize}[topsep=2pt,leftmargin=15pt,itemsep=2pt]
\item {\it Method I:} Define a set of parameters that characterize the QCFs, determine the differential and total cross sections, sample theory-level events from the probability distribution $\hat{\sigma}(x,Q^2)$ over the physical phase space, pass these events through the detector model to obtain synthetic events, use a proper statistical score to quantify the distance between synthetic and measured events, and update the parameters until the distance is minimized.
\item {\it Method II:} Define a set of parameters that characterize the QCFs, determine the differential and total cross sections, pass the probability distribution $\hat{\sigma}(x,Q^2)$ through the detector model (likely a machine-learning-based surrogate model) to obtain $\hat{\sigma}_m(x,Q^2)$ in the measured space, use a proper statistical score to quantify the distance between the continuous function $\hat{\sigma}_m(x,Q^2)$ and the measured events, and update the parameters until this distance is minimized.
\end{itemize}
Either of these methods is repeated until parameters are obtained that give the best agreement between the synthetic events/probability distribution and the measured events. These workflows are illustrated in Fig.~\ref{fig:workflow}.

%---------------------------------------------------------------------
%---------------------------------------------------------------------

The score in Method I must be statistically proper and measure the difference between two probability distributions using only their realizations.
Common choices include the Kolmogorov--Smirnov test~\cite{Massey:1951ks}, Anderson--Darling test~\cite{Anderson:1952ad,Anderson:1954ad}, Maximum Mean Discrepancy~\cite{Gretton:2012mmd}, and Wasserstein distance~\cite{Villani:2009ot}; our preferred choice is the Energy Score~\cite{gneiting2007strictly,Szekely:2013energy,constantinescu2020statistical}. The Energy Score is a strictly proper scoring rule~\cite{gneiting2007strictly} whose underlying energy distance is the rotation-invariant multivariate extension of the classical $L^2$ distance between cumulative distribution functions, and therefore a natural way to lift Cram\'er--von Mises-- and Anderson--Darling--type tests to the multidimensional $(x, Q^2)$ phase space relevant here. 
The score needed in Method II, on the other hand, must measure the difference between two probability distributions when one is represented by a density function and the other by realizations.
While many of the scores used in Method I can be adapted to Method II by expressing them as expectations, a more robust and computationally cheaper choice is the log score~\cite{Pawitan:2013likelihood,thorarinsdottir2013using,gneiting2007strictly}.
The two methods differ in how they treat the comparison between theory and data: Method I naturally operates on events and compares samples to samples, whereas Method II avoids the synthetic-event sampling step by working directly with the theory probability density. Both formulations should formally be equivalent in the limit of a large number of synthetic events. 

In the present study, where the detector is taken to be the identity, Method II is the more direct and computationally efficient choice. There is no detector model to invert, and the loss is evaluated in a single pass over the measured events without pairwise comparisons. We therefore adopt it here, with the loss function given by :
\begin{align}
\mathcal{L}_{\rm log}(\boldsymbol{\theta}) &= \sum_{h =p,n} \bigg[\omega^h\frac{1}{n^h}\sum_{i=1}^{n^h} -\log\big[\hat{\sigma}^h_m(x_i,Q^2_i;\boldsymbol{\theta})\big] \nonumber \allowdisplaybreaks \\
&\hspace*{20mm}
+ \left[[\sigma_{\rm obs}^h - \sigma^h_{\rm tot}(\boldsymbol{\theta})]/\delta \sigma_{\rm obs}^h\right]^2 \bigg],
\label{eq:LS}
\end{align}
where the outer sum runs over the proton and neutron targets, the first term is the log score evaluated using the probability distribution in the measured space [$\hat{\sigma}^h_m$] after passing through the detector model, with the inner sum running over the measured events, and the second term incorporates the total cross sections whose uncertainty is driven by that of the luminosity, which we take here to be 3\%. The hyper-parameter $\omega^h$ is used to facilitate optimization by ensuring that the two terms in the loss function have similar magnitudes, as determined from their typical values after a burn-in phase. Recall, in this initial study we assume a perfect detector, so $\hat{\sigma}^h_m = \hat{\sigma}^h$ of Eq.~\eqref{eq:probdis}.

%-------------------------------------------------------------------------------
%-------------------------------------------------------------------------------
\section{TRADITIONAL HISTOGRAM ANALYSIS FOR DIS\label{sec:binlevel}}
\begin{figure}[tbp]
\centering\includegraphics[width=\linewidth]{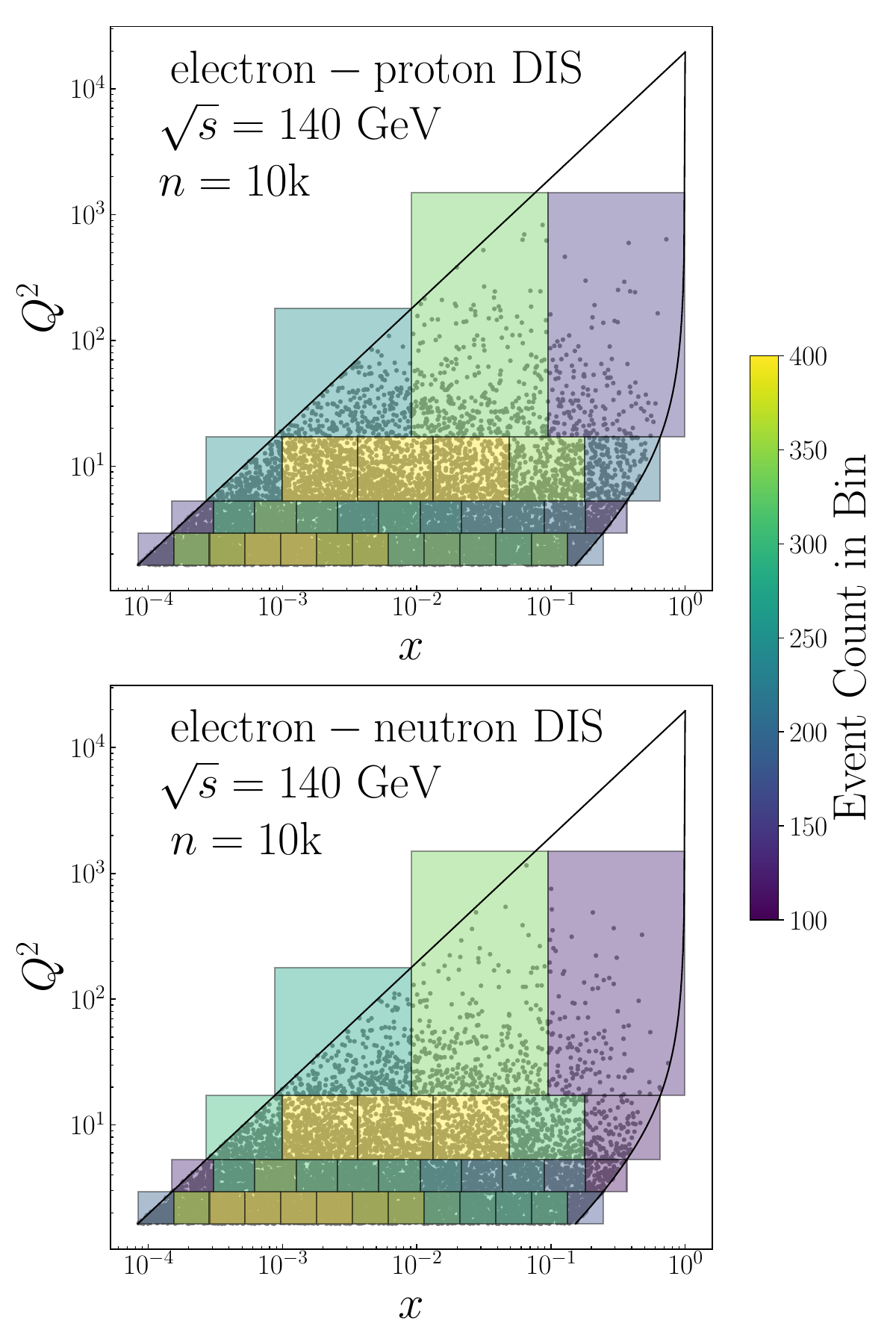}
\caption{Synthetic events for electron-proton (top) and electron-neutron (bottom) deep inelastic scattering at EIC kinematics obtained using the ground truth PDFs. The physical phase space $\Omega$ is given by the domain inside the solid black lines, and a binning of the events is illustrated which is used for the traditional histogram analysis.}
\label{fig:events_histogram}
\end{figure}

To contrast and validate the event-level approach, we also perform a traditional histogram analysis of the same synthetic event and total cross section data. The first step in this process is to bin the unfolded data, so that each bin has sufficient events for meaningful uncertainty quantification. To do this we use adaptive binning, where we start with an oversized number of bins in $x$ (logarithmically spaced) and $Q^2$, and then gradually decrease the number of bins until each bin has at least $1\%$ of the total number of events to ensure approximate statistical Gaussianity. Fig.~\ref{fig:events_histogram} illustrates this adaptive binning for 10,000 synthetic events for electron-proton and electron-neutron scattering at EIC kinematics. The black lines enclose the physical phase space, and as expected the majority of events are at small $x$ and $Q^2$. In both the electron-proton and electron-neutron cases the resulting uncertainties are purely statistical.

To determine the fraction of the total cross section that comes from each bin, we use Eq.~\eqref{eq:eventscrosssection} to obtain 
\begin{align}
\frac{\sigma_{\rm bin}}{\sigma_{\rm obs}} = \frac{n_{\rm bin}}{n_{\rm tot}}, 
\label{eq:binning}
\end{align}
where $n_{\rm bin}$ is the number of events in a bin, $n_{\rm tot}$ is the total number of events observed associated with the physical phase space, $\sigma_{\rm obs}$ is the measured total cross section, and $\sigma_{\rm bin}$ is the fraction of $\sigma_{\rm obs}$ assigned to that bin. Often, $\sigma_{\rm bin}$ is converted to an average differential cross section for that bin, however, this step is not necessary, so our histogram analysis is performed at the $\sigma_{\rm bin}$ level. To compare the histogrammed data with theory (which is determined by a set of parameters $\boldsymbol{\theta}$), we integrate the theory differential cross section over $x$ and $Q^2$ for each bin to determine $\sigma^{\rm thy}_{\rm bin}(\boldsymbol{\theta})$, which is the theory result for the fraction of the total cross section from the physical phase space contained by the bin. For bins on the edges of the physical phase space, we only integrate over the portion of each bin within the physical phase space.

%---------------------------------------------------------------------
%---------------------------------------------------------------------

%---------------------------------------------------------------------
\begin{figure*}[tbp]
\centering\includegraphics[width=\textwidth]{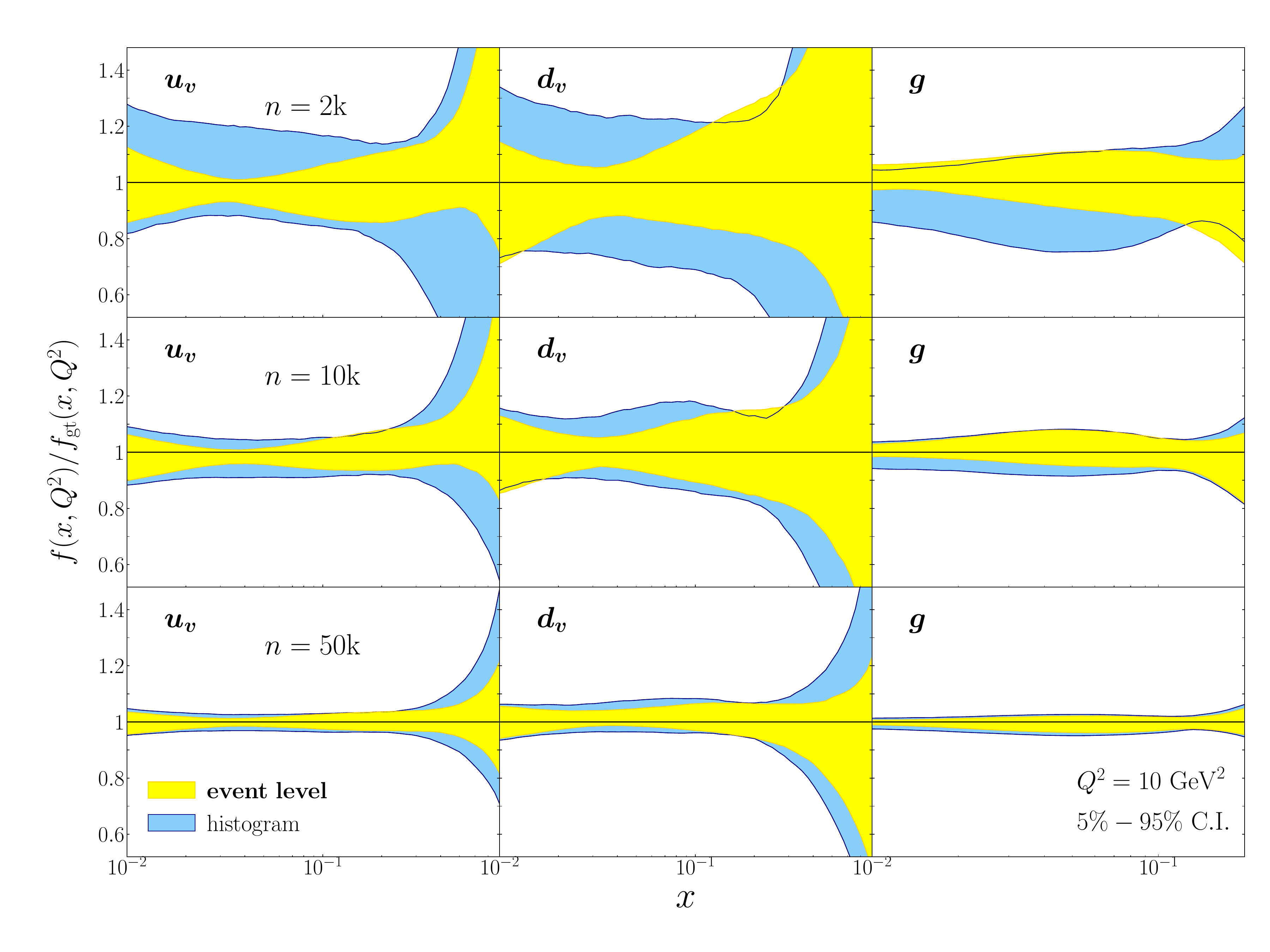}
\caption{Ratio of the extracted $f = u_v,~d_v,~g$ PDFs to the ground truth for the histogram (blue) and event-level (yellow) analyses for $Q^2=10~{\rm GeV}^2$ using 2,000 (top), 10,000 (center), and 50,000 (bottom) events. The uncertainty bands represent a 5-95\% confidence interval.}
\label{f.QCFevents}
\end{figure*}
%---------------------------------------------------------------------

For the histogram-level analysis, statistical uncertainties are assigned to the binned data from the event counts in each bin. We assessed possible bin-to-bin correlations through a bootstrap study in which the original event sample was resampled with replacement and rehistogrammed for each bootstrap replica. The resulting covariance matrix exhibited only small off-diagonal correlations, consistent with residual statistical fluctuations of the bootstrap estimate rather than statistically significant correlations. We therefore neglect bin-to-bin correlations and retain only the diagonal statistical uncertainties. The uncertainty propagation is then performed through a bootstrapping procedure by Gaussianly reshuffling the bin value within its statitistical uncertainty.

In the histogram approach, we minimize a loss function that is defined by the $\chi^2$ according to
\begin{align}
\chi^2(\boldsymbol{\theta}) = \sum_{h = p,n}\ \sum_{{\rm bin}=1}^{n_{\rm bins}}\frac{(\sigma^h_{\rm bin} - \sigma^{{\rm thy},h}_{\rm bin}(\boldsymbol{\theta}))^2}{(\delta^h_{\rm bin})^2},
\label{eq.chi2}
\end{align}
where  $n_{\rm bins}$ is the number of bins in the datasets (not to be confused with individual events), and $\delta^h_{\rm bin}$ is the statistical uncertainty of the integrated cross section within the bin given by $\sigma_{\rm bin}^h/\sqrt{n_{\rm bin}}$, where $n_{\rm bin}$ is the number of events in that bin. Optimizing the overall cross section is not needed for the histogram approach as summing over all the bins will give $\sigma_{\rm obs}$ because of Eq.~\eqref{eq:binning}. In the histogram approach the parameters $\boldsymbol{\theta}$ are tuned to minimize $\chi^2(\boldsymbol{\theta})$.

%-------------------------------------------------------------------------------
%-------------------------------------------------------------------------------
\section{CONSTRASTING EVENT-LEVEL AND HISTOGRAM ANALYSIS FOR DIS\label{sec:results}}
To contrast the event-level and histogram approaches, we perform an analysis using synthetic data for electron-proton and electron-neutron DIS at EIC kinematics. We begin by defining ground truth parameters, $\boldsymbol{\theta}_{\rm gt}$, for the proton and neutron PDFs. These ground truth parameters give PDFs that are consistent with results from Ref.~\cite{Anderson:2024evk}. Using these PDFs, we can construct the ground truth differential cross sections using Eqs.~\eqref{eq:DISxsec}--\eqref{eq:PDFparam} and can integrate these over the physical phase space to obtain the ground truth total cross sections $\sigma_{\rm gt}^{p}$ and $\sigma_{\rm gt}^{n}$. These results can then be used to define the proton and neutron probability distributions given by Eq.~\eqref{eq:probdis}, which are sampled to give two sets of ground truth events in the physical phase space for proton and neutron DIS. For sampling, we use the standard Monte Carlo rejection sampling approach~\cite{devroye_chapter_2006}. These ground truth events then pass through a detector model (which in this case is trivial because we assume a perfect detector) to fold in detector effects to obtain two sets of ground truth events in the measured phase space. 

Using this set of synthetic measured events we can then perform an event-level analysis using the framework discussed in Sec.~\ref{sec:eventlevel} and similarly a histogram analysis using the methods of Sec.~\ref{sec:binlevel}. For this initial study we fit the parameters for the gluon PDFs and the valence up and down quark PDFs, $u_v(x)$ and $d_v(x)$. With constrains from the baryon number and momentum sum rules, this gives 6 free parameters for both the event-level and histogram analyses. 

We perform event-level and histogram inferences of the PDF parameters with 2k, 5k, 10k, and 50k synthetic events for each of electron-proton and electron-neutron DIS. The resulting $u_v$, $d_v$, and $g$ PDFs from our phenomenological analyses, relative to the ground truth, are shown in Fig.~\ref{f.QCFevents} for $n \in \{2{\rm k}, 10 {\rm k}, 50 {\rm k}\}$. The uncertainty bands are the 5\%-95\% confidence interval (C.I.) of the ensemble of replicas. We find that while each of the event level and histogram analyses predict the ground truth within the uncertainty bands, the event-level analysis has noticeably smaller uncertainties, indicating a tighter agreement with the ground truth. Because of the loss of information associated with the binning, we find that bypassing that systematic provides a better prediction for the ground truth. We note that Ref.~\cite{Benato:2025rgo} also found similar tightening of uncertainties in their unbinned analysis relative to their binned one. And as statistically expected, the uncertainties on all distributions decrease with increasing number of events.

To quantify the discrepancies with the ground truth differential cross sections, we devised an integrated weighted mean-squared error metric at a reference scale, $Q^2_{r}$:
\begin{equation}
D^2_h(Q_r^2) = \int_{x_{\rm min}}^{x_{\rm max}} \dd x \left(\frac{\sigma^h(x,Q^2_r;\boldsymbol{\theta}) - \sigma^h(x,Q^2_r;\boldsymbol{\theta}_{\rm gt})}{\sigma^h(x,Q^2_r;\boldsymbol{\theta}_{\rm gt})}\right)^2.
    \label{eq.distance}
\end{equation}
This metric value decreases as the replica more closely resembles the ground truth. We perform a statistical confidence analysis for this metric over the bootstraps. In Tab.~\ref{t.delta}, we show $D = \sqrt{D^2}$ in percentages, to show the integrated difference from the ground truth for each inference at a reference $Q^2_r=10~{\rm GeV}^2$. We show the median values and the confidence level over all of the replicas from the bootstrapped analysis. We find that the event-level analysis out-performs the histogram analyses in all cases in both the median and uncertainties, indicated by a smaller value of $D$. When the number of events increases, the distances get closer together for both the event-level and histogram analyses, implying that the fitted theory better reproduces more precise data and is more constraining in the fit. Qualitatively similar results are found for other $Q^2$ values, with the event level analysis always outperforming the histogram approach.

%---------------------------------------------------------------------
\begin{table}[tbp]
\centering
\addtolength{\tabcolsep}{1.9pt}
\caption{Metric $D$ for $Q_r^2 = 10~\mathrm{GeV}^2$. Uncertainties denote the $5\%$--$95\%$ confidence interval.}
\begin{tabular}{lcccc}
\toprule
 & 2k & 5k & 10k & 50k \\
\midrule
\multicolumn{5}{l}{\textbf{Proton}} \\
Event level 
& $4.40^{+5.10}_{-3.17}\%$ 
& $3.83^{+3.76}_{-2.82}\%$ 
& $2.47^{+3.26}_{-1.86}\%$ 
& $1.01^{+1.26}_{-0.72}\%$ \\[1.0em]
Histogram 
& $8.28^{+12.30}_{-6.14}\%$ 
& $5.52^{+7.64}_{-4.03}\%$ 
& $3.79^{+5.71}_{-2.93}\%$ 
& $1.51^{+2.14}_{-1.09}\%$ \\
\midrule
\multicolumn{5}{l}{\textbf{Neutron}} \\
Event level 
& $6.65^{+9.36}_{-4.70}\%$ 
& $4.89^{+4.82}_{-3.33}\%$ 
& $3.15^{+4.86}_{-2.29}\%$ 
& $1.58^{+2.32}_{-1.18}\%$ \\[1.0em]
Histogram 
& $9.32^{+13.55}_{-6.65}\%$ 
& $6.15^{+7.66}_{-4.10}\%$ 
& $4.23^{+5.88}_{-3.02}\%$ 
& $2.20^{+3.05}_{-1.65}\%$ \\
\bottomrule
\end{tabular}
\label{t.delta}
\end{table}
%---------------------------------------------------------------------

%---------------------------------------------------------------------
\begin{table}[!t]
\centering
\addtolength{\tabcolsep}{7pt}
\caption{Summary of Pearson $\chi^2/n_{\rm bins}$ relative to the histogrammed data from the event-level and histogram analyses.}
\label{t.chi2s}
\begin{tabular}{ccccccc}
\toprule
& \multicolumn{3}{c}{\textbf{Event-level}}
& \multicolumn{3}{c}{\textbf{Histogram}} \\
\cmidrule(lr){2-4}
\cmidrule(lr){5-7}
$N_{\rm evts}$ & $ep$ & $en$ & \textbf{Total}
& $ep$ & $en$ & \textbf{Total} \\
\midrule
\textbf{2k}  & 0.99 & 0.74 & \textbf{0.86} & 0.86 & 0.74 & \textbf{0.80} \\
\textbf{5k}  & 0.57 & 0.72 & \textbf{0.64} & 0.58 & 0.60 & \textbf{0.59} \\
\textbf{10k} & 0.73 & 0.65 & \textbf{0.69} & 0.72 & 0.64 & \textbf{0.68} \\
\textbf{50k} & 0.72 & 0.77 & \textbf{0.75} & 0.72 & 0.77 & \textbf{0.75} \\
\bottomrule
\end{tabular}
\end{table}
%---------------------------------------------------------------------

Another metric to describe the goodness of fit is the Pearson's $\chi^2$ test relative to the histogrammed data. Here, we replace the $\sigma_{\rm bin}^{{\rm thy},h}(\boldsymbol{\theta})$ in Eq.~\ref{eq.chi2} with the expectation value of the theory over all bootstraps, namely, $\mathbb{E}[\sigma_{\rm bin}^{{\rm thy},h}(\boldsymbol{\theta})] = 1/N_{\rm bts}\sum_{i=1}^{n_{\rm bts}} \sigma_{\rm bin}^{{\rm thy},h}(\boldsymbol{\theta}_i)$, where $n_{\rm bts} = 1000$ is the total number of bootstraps  performed. In Tab.~\ref{t.chi2s}, we show the Pearson's $\chi^2/n_{\rm bins}$ metric from each of fits to the event-level and histogram approaches. We observe that each approach has a $\chi^2/n_{\rm bins}$ of less than 1 for all events sets, even though the event-level inference was not performed on the binned events. While a $\chi^2/n_{\rm bins}$ of 1 is the expectation, it is worth noting that other random sets of events from the underlying sampled ground truth follow an expected $\chi^2$ distribution against the ground truth set of parameters. Notably, the event level analysis does not give a smaller $\chi^2$ than the histogram analysis, indicating that the histogram analysis has found its own global minimum given its loss function defined in Eq.~\ref{eq.chi2}. As the number of events increases, the descriptions of the Pearson's $\chi^2$ approaches the same value, indicating a convergence of the event-level and histogram analyses. As the number of events increases, the sampling procedure better represents the underlying ground truth, and both the histogram and event-level analyses provide increasingly similar Pearson's $\chi^2$ distributions.

%-------------------------------------------------------------------------------
%-------------------------------------------------------------------------------
\section{CONCLUSIONS AND OUTLOOK\label{sec:conclusions}}
Using a new event-level analysis framework introduced in this work, we have performed an event level inference on PDFs from synthetic DIS scattering events and total cross sections. This event-level analysis was contrasted with the traditional histogram approach. The event-level framework eliminates the two main sources of information loss in the histogram approach, which are associated with the unfolding and binning procedures. We find that the event-level analysis outperforms the traditional histogram approach in the inference of PDFs. The metric used to demonstrate this is the mean-squared error distance defined in Eq.~\eqref{eq.distance} and summarized in Tab.~\ref{t.delta}. If we consider the proton results with 10,000 events we find that the event-level framework performs around 35\% better than the traditional histogram approach. 

The addition of a detector into the event-level and traditional workflows will result in additional information loss, which is then exacerbated by the unfolding procedure. In future work we will include detector effects and directly contrast the unfolding and folding procedures, where we expect the event-level to further outperform the histogram approach because information loss is minimized by folding in detector effects. Similarly, for differential cross sections associated is 3D quark-gluon imaging such as SIDIS to access TMDs or DVES to access GPDs the challenges associated with binning will be amplified. In these processes, the measured phase space grows with the number of reconstructed final-state particles: SIDIS measures the scattered electron and a fragmenting hadron, and DVES measures the scattered electron, an on-shell photon or hadron, and the recoil target hadron. As a result, the corresponding cross sections depend on several correlated kinematic variables. In practice, measurements are often reported in projected spaces, requiring integrations over the multidimensional theoretical differential cross section. Such projections can reduce sensitivity to correlations among the measured variables and thereby weaken constraints on poorly known QCFs, such as the Sivers function, transversity TMDs, and GPDs. The contrasting of event-level and histogram analysis for 3D imaging is an important future study.

This study was performed with synthetic data, as such, extending this framework to experimentally measured events is essential, especially in light of the growing interest in unbinned and detector-level analyses~\cite{Barrue:2026qul}, and recent efforts to present measurements in event-level formats~\cite{ATLAS:2024xxl}. To this end, we are developing differentiable and surrogate detector models that allow gradients of theory parameters to be propagated through the optimization, and are including theoretical corrections required for detector-level comparisons, such as radiative corrections~\cite{Liu:2020rvc,Liu:2021jfp}.

%...................................................
\begin{acknowledgments}
This work was supported by the U.S.~Department of Energy, Office of Science, Office of Nuclear Physics (NP),  Office of Advanced Scientific Computing Research (ASCR), Scientific Discovery through Advanced Computing (SciDAC) Program through ASCR-NP partnership and the FASTMath Institute under contract no.~DE-AC02-06CH11357.
\end{acknowledgments}

%-------------------------------------------------------------------------------
%-------------------------------------------------------------------------------
\makeatletter
\def\@eprint#1{}
\makeatother

\bibliography{bibliography}

\end{document}

%% file: texdefinitions_papers.tex
\usepackage[utf8]{inputenc}
\usepackage{newtxtext}
\usepackage[libertine]{newtxmath}
\usepackage[main=english]{babel}
\usepackage{microtype}

\usepackage{graphicx}
\usepackage[usenames,dvipsnames]{xcolor}
\graphicspath{{figures/}}

\usepackage{hyperref}
\hypersetup{colorlinks=true,citecolor=blue,linkcolor=blue,urlcolor=blue}

\usepackage{diagbox}
\usepackage{booktabs}

\usepackage{amsmath,amsfonts}
\usepackage{mathrsfs}
\usepackage{slashed}
\usepackage{bm}

\usepackage{outlines}
\usepackage{enumitem}
\setenumerate[1]{label=\Roman*.}
\setenumerate[2]{label=\Alph*.}
\setenumerate[3]{label=\roman*.}
\setenumerate[4]{label=\alph*.}

\usepackage[compact]{titlesec}
\newcommand*\dd{\mathop{}\!\mathrm{d}}

\newcommand*\emdash{\hspace{1pt}---\hspace{1pt}}
\newcommand*\lf{\left}
\newcommand*\rg{\right}

\usepackage[normalem]{ulem}
\usepackage{colortbl}